\documentclass[fleqn,twocolumn,10pt]{wlscirep}

\usepackage{setspace}
\usepackage[english]{babel}
\usepackage[utf8]{inputenc}
\usepackage{amsmath,amsfonts,amsthm,bm}
\usepackage{graphicx}
\usepackage[]{float}
\usepackage{caption}
\usepackage{subcaption}
\usepackage{textcomp}
\usepackage{authblk}
\usepackage{color}
\usepackage{amsmath}
\usepackage{gensymb}
\usepackage{braket}
\usepackage{comment}
\usepackage{color,soul}
\usepackage{bm}

\usepackage[T1]{fontenc}
\usepackage[symbol*]{footmisc}
\usepackage[numbers,sort&compress]{natbib}
\usepackage{xcolor}

\usepackage{subfiles}

\title{High-fidelity spatial mode transmission through a 1-km-long multimode fiber via vectorial time reversal} 

\author[1,*]{Yiyu Zhou}
\author[2]{Boris Braverman}
\author[3]{Alexander Fyffe}
\author[4]{Runzhou Zhang}
\author[1]{Jiapeng Zhao}
\author[4]{Alan E. Willner}
\author[3]{Zhimin Shi}
\author[1,2]{Robert W. Boyd}
\affil[1]{The Institute of Optics, University of Rochester, Rochester, New York 14627, USA}
\affil[2]{Department of Physics, University of Ottawa, Ottawa, Ontario K1N 6N5, Canada}
\affil[3]{Department of Physics, University of South Florida, Tampa, Florida 33620, USA}
\affil[4]{Department of Electrical and Computer Engineering, University of Southern California, Los Angeles, California, 90089, USA}
\vspace{5mm}
\affil[*]{Corresponding author: yzhou62@ur.rochester.edu}

\begin{abstract}
\textbf{The large number of spatial modes supported by standard multimode fibers is a promising platform for boosting the channel capacity of quantum and classical communications by orders of magnitude. However, the practical use of long multimode fibers is severely hampered by modal crosstalk and polarization mixing. To overcome these challenges, we develop and experimentally demonstrate a vectorial time reversal technique, which is accomplished by digitally pre-shaping the wavefront and polarization of the forward-propagating signal beam to be the phase conjugate of an auxiliary, backward-propagating probe beam. Here, we report an average modal fidelity above 80\% for 210 Laguerre-Gauss and Hermite-Gauss modes by using vectorial time reversal over an unstabilized 1-km-long fiber. We also propose a practical and scalable spatial-mode-multiplexed quantum communication protocol over long multimode fibers to illustrate potential applications that can be enabled by our technique.}
\end{abstract}

\begin{document}

\maketitle


\pagestyle{plain}

\subsection*{Introduction}
Quantum cryptography \cite{gisin2002quantum, scarani2009security} is a maturing technology that can guarantee the security of communication based on the fundamental laws of physics. The secure key rate of quantum key distribution (QKD) systems is many orders of magnitude lower than the data transfer rate of classical communication systems, inhibiting the widespread adoption of QKD in practical scenarios. Numerous methods have been proposed to increase the secure key rate in QKD, such as development of new protocols \cite{lucamarini2018overcoming}{}, use of high-performance detectors \cite{zhang2015advances}{}, and wavelength-division multiplexing \cite{patel2014quantum}{}. The spatial degree of freedom in a multimode fiber (MMF) has long been recognized as an additional resource to further increase the communication rate by either mode-division multiplexing \cite{gibson2004free,wang2012terabit,bozinovic2013terabit,liu2018direct,flaes2018robustness,luo2014wdm,gregg2015conservation, xavier2020quantum}{} or high-dimensional encoding \cite{mirhosseini2015high,mafu2013higher,vallone2014free,zhou2019using, wang2020high}{}. It is compatible with other multiplexing methods such as wavelength-division multiplexing and can be also used to enhance quantum teleportation \cite{yin2012quantum}{} and entanglement distribution \cite{loffler2011fiber,liu2020multidimensional}{}. However, the inevitable mode crosstalk in standard MMFs is a persistent obstacle to practical applications of spatial modes for QKD. Tremendous efforts have been devoted to attempts to mitigate the effects of spatial mode crosstalk during the past decades.
Transfer matrix inversion is a standard method that has been successfully used to transmit spatial modes through MMFs \cite{carpenter2014110x110,ploschner2015seeing,gordon2019characterising,mounaix2019time, mounaix2019control, xiong2018complete, xiong2019long}{}. However, standard MMFs can support between tens and hundreds of modes depending on the wavelength, and thus the number of complex-valued elements in the transfer matrix is typically between $10^3$ and $10^5$. As a consequence, all transfer matrix inversion experiments reported in the literature have used a short MMF ($\approx$1~m) \cite{carpenter2014110x110,ploschner2015seeing,gordon2019characterising,mounaix2019time, mounaix2019control, xiong2018complete, xiong2019long}{} because the fiber has to be carefully stabilized during the slow characterization process (see Supplementary Note 1 for a summary of the fiber length used in previously reported experiments). When applying this method to a long fiber, it is foreseeable that instability will severely impede long-distance communication outside the laboratory. By contrast, mode-group excitation \cite{ryf2015mode,franz2012experimental,zhu2017orbital}{} has been applied to long fibers due to the relatively low inter-modal-group crosstalk. However, for a fiber supporting $N$ spatial modes, only approximately $\sqrt{N}$ mode groups are supported. Thus the number of usable mode groups is intrinsically limited in this method (see Supplementary Note 1). Multiple-input-multiple-output (MIMO) algorithm is another standard method for classical crosstalk mitigation \cite{amphawan2011review}{}. However, it requires a high signal-to-noise ratio for digital signal processing and thus is unsuitable for quantum applications. Hence, none of these existing methods can be used to support a large number ($>100$) of modes for high-dimensional or spatial-mode-multiplexed QKD over long, unstabilized, standard MMFs outside the laboratory.

\begin{figure}[t]
\center
\includegraphics[width= \linewidth]{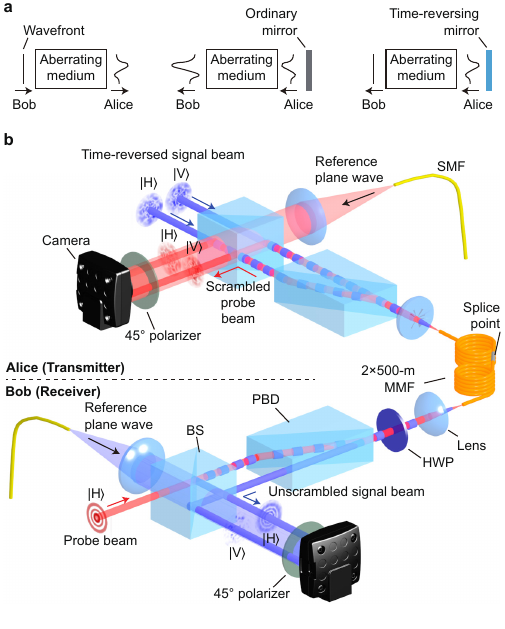}
\caption{\textbf{Illustration of experiment.} \textbf{a} The concept of time reversal. \textbf{b} Schematic of the experiment to send high-fidelity spatial modes from Alice to Bob. Bob first transmits a probe beam to Alice (denoted by red beams). Alice performs vectorial off-axis holography on her received probe beam and generates the corresponding time-reversed signal beam (denoted by blue beams). $\ket{\text{H}}$ and $\ket{\text{V}}$ stand for the horizontal and vertical polarization state respectively. HWP: half-wave plate. PBD: polarizing beam displacer. SMF: single-mode fiber. BS: beamsplitter. See Supplementary Note 2 for experimental details.}
\label{fig:setup}
\end{figure}

Optical time reversal, which is also referred to as phase conjugation \cite{boyd2003nonlinear}{}, is an alternative method for modal crosstalk suppression. The concept of time reversal is illustrated in Fig.~\ref{fig:setup}(a). The wavefront of an optical beam transmitted by Bob is distorted by an aberrating medium as shown in the left panel. Reflecting the beam by an ordinary mirror at Alice's side is not helpful as the wavefront becomes distorted even more severely (see the middle panel). In contrast, a time-reversing mirror flips the sign of phase of the reflected beam, and consequently wavefront distortion can be exactly corrected after propagating through the same aberrating medium \cite{boyd2003nonlinear}{} as illustrated in the right panel. Although we use a simple plane wave to illustrate the concept, it should be noted that time reversal is applicable to an arbitrary spatial mode. Optical time reversal has been investigated for biological tissues \cite{yaqoob2008optical, wang2015focusing, yang2019angular, shen2016focusing,judkewitz2013speckle, mosk2012controlling, cui2010implementation,  papadopoulos2012focusing, ma2018reconstruction,ma2020structured, morales2015delivery}{} and MMFs \cite{ma2020structured, ma2018reconstruction, morales2015delivery, papadopoulos2012focusing, czarske2016transmission, bae2019compensation}{}. The textbook description of time reversal \cite{boyd2003nonlinear}{} assumes a scalar incident field $\textbf{\text{E}}_{\text{scalar}}(\textbf{\text{r}}) = \hat{\epsilon} A(\textbf{\text{r}}) e^{i \textbf{\text{k}} \cdot \textbf{\text{r}}}$, where $\hat{\epsilon}$ is the polarization unit vector, $A(\textbf{\text{r}}) $ is the complex field amplitude, $\textbf{\text{r}}$ is the position vector, and $\textbf{\text{k}}$ is the wavevector of the incident field. It can be seen that the polarization $\hat{\epsilon} $ and the complex field amplitude $A(\textbf{\text{r}}) $ are separable from each other. The scalar time reversal of the incident field can be expressed as \cite{boyd2003nonlinear}{}
\begin{equation}
\textbf{\text{E}}_{\text{scalar}}^*(\textbf{\text{r}}) =  \hat{\epsilon}^* A^*(\textbf{\text{r}}) e^{-i \textbf{\text{k}} \cdot \textbf{\text{r}}},
\label{eq:convTR}
\end{equation}
where * in the superscript denotes the complex conjugate. Here we generalize the scalar time reversal and propose the vectorial time reversal. Assume a vectorial incident field given by $\textbf{\text{E}}_{\text{vector}}(\textbf{\text{r}}) = \hat{x} A_1(\textbf{\text{r}}) e^{i \textbf{\text{k}}_1 \cdot \textbf{\text{r}}} + \hat{y} A_2(\textbf{\text{r}}) e^{i \textbf{\text{k}}_2 \cdot \textbf{\text{r}}}$, where $A_1(\textbf{\text{r}})$ and $A_2(\textbf{\text{r}})$ ($\textbf{\text{k}}_1$ and $\textbf{\text{k}}_2$) denote the respective complex field amplitude (wavevector) of the horizontally and vertically polarized field, and $\hat{x}$ ($\hat{y}$) denotes the horizontal (vertical) unit vector. Its vectorial time reversal can be written as
\begin{equation}
\textbf{\text{E}}_{\text{vector}}^*(\textbf{\text{r}}) =  \hat{x} A_1^*(\textbf{\text{r}}) e^{-i \textbf{\text{k}}_1 \cdot \textbf{\text{r}}} + \hat{y} A_2^*(\textbf{\text{r}}) e^{-i \textbf{\text{k}}_2 \cdot \textbf{\text{r}}},
\label{eq:vecTR}
\end{equation}
Here, the incident vectorial field is described by a non-separable state, which is a more general form than a separable state. All experimental demonstrations in MMFs \cite{ma2020structured, ma2018reconstruction, morales2015delivery, papadopoulos2012focusing, czarske2016transmission, bae2019compensation}{} to date have been solely based upon scalar time reversal. However, the scalar time reversal can only succeed in a short MMF ($\approx$1~m) \cite{ma2020structured, ma2018reconstruction, morales2015delivery, papadopoulos2012focusing, czarske2016transmission}{} or a few-mode fiber \cite{bae2019compensation}{}, because the optical field scrambled by a long MMF has to be described by a vectorial field due to the spatially-varying birefringence in a MMF.

In this work, we experimentally show how vectorial time reversal can be used to support 210 modes over a 1-km-long MMF. The fiber length used in our experiment is nearly three orders of magnitude greater than those used in many previously reported experiments demonstrating high-fidelity spatial mode transmission through MMFs (see Supplementary Note 1). To illustrate the potential of our method, a spatial-mode-multiplexed QKD protocol is proposed that is suitable for realization in real-world, unstable links with a practical, scalable implementation.

\subsection*{Results}
\noindent\textbf{Schematic of vectorial time reversal.} Figure \ref{fig:setup}(b) presents the conceptual schematic of our experiment. Here we experimentally demonstrate that digital vectorial time reversal can be successfully applied to transmit 210 high-fidelity spatial modes (up to mode group 13) through a 1-km-long, standard, graded-index MMF with the number of used modes limited by the active area size of our SLM. We choose the commonly used Laguerre-Gauss and Hermite-Gauss modes for demonstration because they are the eigenmodes of a graded-index MMF \cite{mafi2012pulse}{}, which exhibit better robustness and minimized loss during propagation \cite{flaes2018robustness}{} compared to other basis sets and thus enable time reversal to a full extent. A 780~nm laser is used as the light source in the experiment. Bob prepares a spatial mode of interest (i.e., a probe beam) and transmits it to Alice through a 1-km-long, standard, graded-index MMF (Clearcurve OM3, Corning). The fiber is comprised of two 500-m-long bare fibers that are spliced together and are free of any specialized thermal or mechanical isolation. The fiber has a core diameter of 50~\textmu{}m and ${\rm{NA}}=0.2$, therefore supporting $\approx$400 modes per polarization at 780~nm. The polarization of the probe beam transmitted by Bob can be adjusted by a half-wave plate. After transmission through the fiber, the scrambled probe beam received by Alice has a distorted spatial and polarization profile. Alice then performs vectorial off-axis holography to measure the spatial and polarization profile of the scrambled probe beam. First, Alice uses a polarizing beam displacer to coherently separate the horizontally and vertically polarized components of the scrambled probe beam into two beams that propagate along the same direction but are transversely displaced with respect to each other. These two beams are then combined with a coherent, 45\textdegree{} polarized reference plane wave at a beamsplitter, and the resultant interference pattern is recorded by a camera. Through off-axis holography \cite{cuche2000spatial}{}, the amplitude, phase, and polarization of the scrambled probe beam can be simultaneously determined via a single-shot measurement \cite{zhu2019single}{} (see Supplementary Note 2). Alice then uses a single spatial light modulator (SLM) to generate the back-propagating signal beams, which are the phase conjugate of the displaced, scrambled probe beams. The two back-propagating time-reversed signal beams are combined coherently by the same polarizing beam displacer. After passing through the same MMF, the signal beam is unscrambled and becomes the mode originally transmitted by Bob with a reversed wavefront. Vectorial off-axis holography is then performed by Bob to quantitatively characterize the spatial and polarization profile of the unscrambled signal beam. Additional experimental details are provided in Supplementary Note 2.

\begin{figure}[t]
\center
\includegraphics[width= \linewidth]{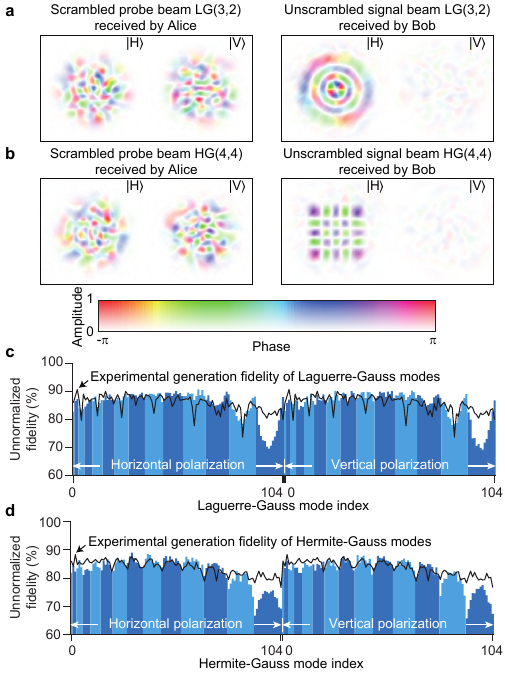}
\caption{\textbf{Modal fidelity measurement.} \textbf{a}, \textbf{b} The measured amplitude, phase, and polarization of the scrambled probe beams and unscrambled signal beams for horizontally polarized LG(3,2) and HG(4,4) mode respectively. \textbf{c}, \textbf{d} The unnormalized modal fidelity for unscrambled Laguerre-Gauss modes and Hermite-Gauss modes. A single index is used to denote the two mode indices for simplicity (see Supplementary Note 3), and the light and dark blue bands denote the odd and even mode group number respectively.}
\label{fig:crosstalk}
\end{figure}

\begin{figure}[t]
\center
\includegraphics[width= \linewidth]{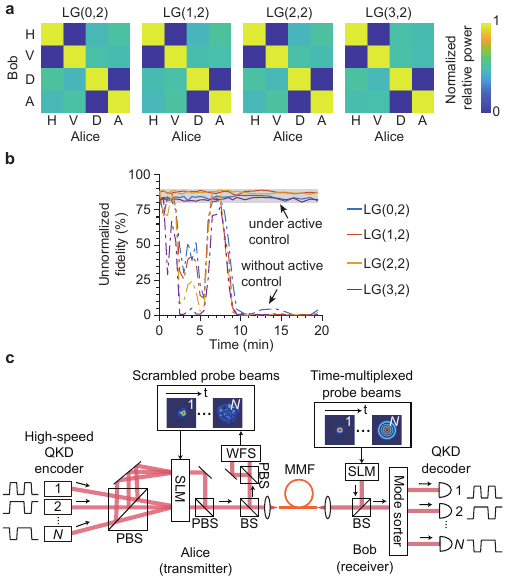}
\caption{\textbf{System performance evaluation.} \textbf{a} The normalized crosstalk matrix in the polarization subspace for horizontal (H), vertical (V), diagonal (D), and anti-diagonal (A) polarizations for LG(0,2), LG(1,2), LG(2,2), and LG(3,2) modes. \textbf{b} Stability test for vectorial time reversal. The unnormalized modal fidelity is measured as a function of time. The shaded area corresponds to modal fidelity between 80\% and 90\%. The solid lines represent the results when the SLM is under active control and the dashed lines represent the results without active control. \textbf{c} Proposed spatial-mode-multiplexed QKD protocol. A densely encoded computer-generated hologram imprinted on a single SLM can be used to simultaneously generate and multiplex a large number of spatial modes. BS: beamsplitter.}
\label{fig:QKD}
\end{figure}

\noindent\textbf{Characterization of modal fidelity.} Figure~\ref{fig:crosstalk}(a, b) shows two examples of experimentally measured scrambled probe beams received by Alice and the unscrambled signal beams received by Bob for horizontally polarized LG(3,2) and HG(4,4) modes, where the mode indices of Laguerre-Gauss mode are denoted by LG($p$,$\ell$) and that of Hermite-Gauss mode are denoted by HG($m$,$n$). It can be seen that the scrambled probe beams are vectorial fields and cannot be described by a separable state, because field profiles of horizontal and vertical polarization are very different. For each unscrambled signal beam received by Bob, we digitally project the mode to an orthonormal spatial mode basis set to calculate the crosstalk matrix. We measure the crosstalk matrix for 105 Laguerre-Gauss modes with $2p+|\ell| \leq 13$ in both horizontal and vertical polarization basis sets, resulting in a 210$\times$210 crosstalk matrix. The same measurement is also performed for Hermite-Gauss modes with $m+n\leq 13$. The unnormalized modal fidelity for individual spatial modes ($i.e.$, the diagonal elements of crosstalk matrix) is shown in Fig.~\ref{fig:crosstalk}(c) with an average of 85.6\% for Laguerre-Gauss modes and Fig.~\ref{fig:crosstalk}(d) with an average of 82.6\% for Hermite-Gauss modes. Here the crosstalk matrix element is calculated as $M_{k',k}=|\braket{\phi_{k'}^{\text{ideal}}|\phi_k^{\text{exp}}}|^2$ and the unnormalized modal fidelity is $F_k=M_{k,k}$, where $\ket{\phi_{k'}^{\text{ideal}}}$ is the ideal spatial mode with a mode index of $k'$, $\ket{\phi_k^{\text{exp}}}$ is the experimentally measured spatial mode with a mode index of $k$, $\braket{\phi_{k'}^{\text{ideal}}|\phi_{k'}^{\text{ideal}}}=1$ and $\braket{\phi_k^{\text{exp}}|\phi_k^{\text{exp}}}=1$. The normalized modal fidelity within the 210-mode subspace has an average of 91.5\% for Laguerre-Gauss modes and 89.3\% for Hermite-Gauss modes, where the normalized modal fidelity is defined as $F_k^{\text{norm}}=M_{k,k} / \sum_{k'=0}^{209} M_{k',k} $. The full 210$\times$210 crosstalk matrices for both Laguerre-Gauss and Hermite-Gauss modes are provided in Supplementary Note 4. Although the performance is characterized using a classical light source and a classical detector, our method is readily applicable to QKD by attenuating the light intensity to a single-photon level and by using single-photon detectors \cite{lo2005decoy}{}. The additional noise of single-photon detectors is outside the scope of this work. Here we attribute the imperfect modal fidelity mainly to the imperfect mode generation by the SLM. To test this hypothesis, we experimentally characterize the fidelity of the probe beam generated by Bob and that of the signal beam generated by Alice. The product of these two fidelities is referred to as experimental generation fidelity, which is presented as solid lines in Fig.~\ref{fig:crosstalk}(c, d) for individual spatial modes (see Supplementary Note 5 for details). It can be seen that the fidelity of unscrambled signal beams received by Bob matches well with the experimental generation fidelity. Hence, the fidelity of unscrambled signal beams is limited by our experimental apparatus (e.g., the SLM) rather than the fiber length, and we believe that our method can be applicable to even longer fibers. We also note that such high modal fidelity is exclusively enabled by vectorial time reversal, while scalar time reversal can only achieve an average unnormalized modal fidelity of 41.2\% for Laguerre-Gauss modes and 39.7\% for Hermite-Gauss modes (see Supplementary Note 6). Nonetheless, there exists a deviation between the unscrambled modal fidelity and the experimental generation fidelity for high-order modes, which we attribute to the fact that high-order modes are susceptible to mode-dependent loss induced by fiber bending and splicing. Further studies are needed to identify the exact reason in order to further expand the transmission distance for high-order modes. To evaluate the performance of our system for polarization-based QKD, we measure the 4$\times$4 polarization crosstalk matrix for each mode within the corresponding spatial mode subspace. The resultant normalized polarization crosstalk matrices for LG(0,2), LG(1,2), LG(2,2), and LG(3,2) are shown in Fig.~\ref{fig:QKD}(a). The average polarization crosstalk is 0.04\% for Laguerre-Gauss modes and 0.05\% for Hermite-Gauss modes (see Supplementary Note 7), which suggests that both the spatial mode and polarization scrambling can be well suppressed through vectorial time reversal. These high-fidelity results directly indicate that the polarization-based QKD protocol can be performed through MMFs, and the secure key rate can be significantly boosted by mode-division multiplexing. Furthermore, because these high-fidelity results are obtained in a spliced fiber, we expect that the vectorial time reversal would also be realized in a much longer fiber. We also calculate the crosstalk matrix for the scrambled probe beams received by Alice in the absence of vectorial time reversal, and the average unnormalized modal fidelity in this case is $\approx$1\% for both Laguerre-Gauss modes and Hermite-Gaussian modes (see Supplementary Note 4), which shows the strong mode scrambling in fiber and by contrast highlights the effectiveness of our method.

\noindent\textbf{Response time of the vectorial time reversal system.} To overcome environmental instability, which is an inevitable concern for long fibers in a real-world environment, vectorial off-axis holography needs to be performed repeatedly in real time, and the phase pattern on Alice's SLM should be updated accordingly to compensate for instability. In the following, we evaluate the response time of our vectorial time reversal system. To perform off-axis holography, we need to retrieve a single-shot image from the camera (which takes 2.6~ms) and execute fast Fourier transforms and interpolations as digital signal processing (which takes 18~ms on a desktop computer, see Supplementary Note 2). It should be noted that the data processing time can be significantly reduced by using a dedicated digital signal processor or even eliminated by careful experimental design and alignment as discussed in Supplementary Note 2. The response time of our system is therefore only constrained by the refresh rate of SLM, which is 4~Hz in our experiment. However, this constraint can be readily removed by using a commercially available fast digital micromirror device (above 10 kHz refresh rate \cite{takaki2009hologram}{}) or a high-speed SLM (sub kHz refresh rate \cite{chen2018pursuing}{}). In addition, we emphasize that the data transfer rate using each spatial mode is not limited by the refresh rate of SLM or the response time of the time-reversal system but is determined by the modulation rate of the modulator used by the encoder \cite{chen2016demonstration}{}, and the time-reversal system response time only needs to be faster than the environmental fluctuation rate in order to overcome instability. In this proof-of-principle experiment, we implement digital vectorial time reversal for one spatial mode at a time, but we emphasize that our method can be readily used to enable mode-division multiplexing \cite{bae2019compensation}{}: by separately pre-shaping individual wavefronts of multiple high-speed-modulated signal beams, high-fidelity spatial modes can be recovered at the receiver, and thus the data streams can be demultiplexed with low crosstalk. To test the operation stability of our system, we also measure the unnormalized modal fidelity as a function of time while the SLM is actively updated every $\approx$30 seconds for each mode, which is depicted by the solid lines in Fig.~\ref{fig:QKD}(b); here the dashed lines represent the modal fidelity in the absence of active control of the SLM. The autocorrelation $R(\Delta t) = \langle | \braket{\phi(t)|\phi(t+\Delta t)} |^2  \rangle$ is calculated according to these data, where $\ket{\phi(t)}$ is the time-reversed mode at time $t$, $\langle \cdot \rangle$ is the time average, and $\ket{\phi(t)}$ is normalized such that $\braket{\phi(t) | \phi(t)}=1$. The time for $R(\Delta t)$ to drop to $1/e$ is approximately 120~s for LG(0,2) and 100~s for LG(3,2). These results clearly show that our system is able to overcome environmental instability even though the unprotected 1-km-long bare fiber is placed on an optical table that has not been floated and is free of any thermal or mechanical isolation. The image of the unprotected fiber spool is shown in Supplementary Note 2. We believe that by using a fast SLM, real-time crosstalk suppression can be achieved even in a harsh environment through a much longer fiber.

\noindent\textbf{Proposed quantum communication protocol.} Figure \ref{fig:QKD}(c) presents a practical, scalable QKD protocol with mode-division multiplexing. Each spatial mode can be used as an independent channel, with time-bin encoding \cite{islam2017provably, boaron2018secure, yin2016measurement, jin2019genuine}{} or continuous-variable encoding \cite{jouguet2013experimental, pirandola2015high, weedbrook2012gaussian}{} used to guarantee communication security within each channel. This spatial-mode-multiplexed setup can also be used to implement high-dimensional encoding by transmitting one spatial mode at a time. The Laguerre-Gauss and Hermite-Gauss modes can be employed as mutually partially unbiased bases for high-dimensional encoding to guarantee security \cite{wang2020high}{}. In particular, it has been previously demonstrated that a single SLM can be used to simultaneously generate and multiplex 105 spatial modes by using a densely encoded computer-generated hologram \cite{trichili2016optical}{}. To implement the $N$-mode-multiplexed QKD protocol, Bob first sequentially transmits $N$ probe beams of interest to Alice through the fiber via spatial mode switching, which can be realized by an acousto-optic modulator with a mode switching rate up to 500~kHz \cite{braverman2020fast}{}. Alice measures the scrambled probe beams using a wavefront sensor (WFS), computes the densely encoded hologram, and then imprints the hologram onto her SLM. Although the densely encoded hologram has a low efficiency, this will not reduce the secure key rate of the attenuated-coherent-state-based QKD because strong loss is inherently necessary to attenuate a classical laser to the single-photon level. The WFS can be realized by either the off-axis holography presented in this work or alternative methods such as the commercial Shack-Hartmann WFS and the vectorial complex field direct measurement \cite{zhu2019single}{}. Alice then prepares $N$ attenuated signal beams with high-speed time-bin encoding or continuous-variable encoding and illuminates the SLM with these beams incident at different angles. These $N$ beams can be obtained by using a single laser with a 1-to-$N$ fiber beamsplitter. The loss induced by the beamsplitter at the transmitter's side is not a concern for attenuated-coherent-state-based QKD protocols. The horizontal and the vertical polarization components are split by a polarizing beamsplitter (PBS) and are incident at two separation locations on the SLM, which allows for generation of vectorial time reversal. Alice's SLM converts each of the $N$ incident beams into the phase conjugate of its corresponding scrambled probe beam, in addition to multiplexing all the modes to propagate in the same direction. The horizontal and vertical polarization components are recombined by another PBS and finally transmitted to Bob through the MMF. As a consequence, all channels can have a high-fidelity spatial profile at Bob's side, and Bob can use the well-developed Laguerre-Gauss or Hermite-Gauss mode sorter \cite{zhou2017sorting,zhou2018hermite, ruffato2018compact}{} to demultiplex the signal beams.

\subsection*{Discussion}

The loss induced by the beamsplitter at the receiver's side can be reduced by using a high-transmission beamsplitter (such as a T:R=95:5 beamsplitter). We note that the reduced secure key rate caused by this small amount of additional loss can be well compromised by the capacity improvement of mode-division multiplexing. To overcome environmental instability, Bob needs to periodically send probe beams of interest to Alice, who updates the phase pattern on the SLM accordingly. The polarization degree of freedom can also be included to further increase the channel capacity. It should be noted that the signal transfer speed is determined by the QKD encoder, not the SLM refresh rate. We emphasize that an analogous protocol, to the best of our knowledge, cannot be realized in a straightforward manner by any alternative methods for the following reasons. Since a complete knowledge of the complex-valued transfer matrix is not needed, our method can be applied to unstabilized, long MMFs outside the laboratory, which is not possible by slow, conventional transfer matrix inversion. MIMO is not applicable to QKD because it requires a large number of photons for digital signal processing. Mode-group excitation only allows for a small number ($\approx$10) of mode groups and is thus unable to fully utilize the channel capacity of the link. Thus, vectorial time reversal offers a unique and practical approach towards spatial-mode-multiplexed quantum communication over realistic, unstable links.

In summary, we have demonstrated that, through the use of vectorial time reversal, we can establish a high-fidelity, 1-km-long communication link that supports 210 spatial modes of a standard MMF. Both spatial mode crosstalk and polarization scrambling in MMF can be well suppressed, which demonstrates the possibility of boosting the communication rate of both classical communication and QKD by either mode-division multiplexing or high-dimensional encoding. In particular, we propose a spatial-mode-multiplexed QKD protocol and show how our method can be used to boost the channel capacity in a straightforward manner. While specialty fibers such as multi-core fibers \cite{bacco2019boosting}{} can also be used to realize space-multiplexed QKD, standard MMFs are cheaper and already widely deployed in existing commercial fiber communication systems, and thus the capacity of these systems can be readily improved by two orders of magnitude using our method without replacing the fibers. Our results also confirm the time reversal symmetry of beam propagation over a 1-km-long MMF. Given the scalability of the experimental implementation and high fidelity of the data, our technique can be useful to not only mode-division multiplexing but also to other applications such as fiber endoscopy \cite{turtaev2018high}{}, lensless microscopy \cite{vcivzmar2012exploiting}{}, and high-dimensional entanglement distribution \cite{loffler2011fiber,liu2020multidimensional}{}.

\section*{Acknowledgements}
This work is supported by the U.S. Office of Naval Research (N00014-17-1-2443, N00014-20-1-2558, N00014-16-1-2813). B.B. acknowledges the support of the Banting Postdoctoral Fellowship. R.W.B. acknowledges funding from the Natural Sciences and Engineering Research Council of Canada, the Canada Research Chairs program, and the Canada First Research Excellence Fund. A.E.W. acknowledges the support of the Vannevar Bush Faculty Fellowship sponsored by the Basic Research Office of the Assistant Secretary of Defense for Research and Engineering. R.Z. acknowledges the support of the Qualcomm Innovation Fellowship.

\section*{Competing interests}
The authors declare no competing interests.

\section*{Author contributions}
Y.Z. conceived and performed the experiment with assistance from B.B., Z.S., R.Z., J.Z., and R.W.B. Y.Z., B.B., A.F., R.Z., J.Z., A.E.W., Z.S., and R.W.B. contributed to the discussion of the results and the writing of the manuscript. A.E.W., Z.S., and R.W.B. supervised the project.

\section*{Data availability}
The data that support the findings of this study are available from the corresponding author upon reasonable request.

\section*{Code availability}
All relevant computer codes supporting this study are available from the corresponding author upon reasonable request.


\onecolumn
\clearpage
\begin{center}
\Large{\textbf{Supplementary Information}}
\end{center}

\setstretch{1.25}
\setcounter{equation}{0} \setcounter{subsection}{0} \setcounter{section}{0}
\setcounter{figure}{0}
\renewcommand{\theequation}{S\arabic{equation}}
\renewcommand{\refname}{Supplemental References}
\renewcommand{\bibnumfmt}[1]{[S#1]} 
\renewcommand{\citenumfont}[1]{S#1}

\renewcommand{\figurename}{Supplementary Figure}
\renewcommand{\tablename}{Supplementary Table}
\renewcommand{\thesection}{Supplementary Note \arabic{section}}
\renewcommand{\thesubsection}{\thesection.\arabic{subsection}}
\renewcommand{\thesubsubsection}{\thesubsection.\arabic{subsubsection}}

\section{--- Literature review}
\begin{table}[h]
\centering
\begin{tabular}{|c|c|c|c|c|}
\hline
No. & Method & \begin{tabular}[c]{@{}c@{}}Multimode \\ fiber length\end{tabular} & \begin{tabular}[c]{@{}c@{}}Number of \\ demonstrated modes\end{tabular} &  Reference \\ \hline
1 & Vectorial time reversal & 1 km & 210 HG/LG modes  & This work \\ \hline
2 & Transfer matrix inversion & 2 m & 6 discrete spot modes  & \cite{valencia2020unscrambling}{} \\ \hline
3 & Transfer matrix inversion & 5 m & N/A & \cite{mounaix2020time}{} \\ \hline
4 & Transfer matrix inversion & 2 m & 110 LP modes & \cite{carpenter2014110x110}{} \\ \hline
5 & Transfer matrix inversion & 0.3 m & 110 LP modes  & \cite{ploschner2015seeing}{} \\ \hline
6 & Transfer matrix inversion & 0.5 m & 18 discrete spot modes &  \cite{leedumrongwatthanakun2020programmable}{} \\ \hline
7 & Transfer matrix inversion & 2 m & N/A &  \cite{gordon2019characterizing}{} \\ \hline
8 & Transfer matrix inversion & 1 m & N/A &  \cite{mounaix2019control}{} \\ \hline
9 & Transfer matrix inversion & 2 m & N/A &  \cite{xiong2018complete}{} \\ \hline
10 & Transfer matrix inversion & 1 m & N/A &  \cite{xiong2019long}{} \\ \hline
11 & Scalar time reversal & 1 m & N/A &  \cite{papadopoulos2012focusing}{} \\ \hline
12 & Scalar time reversal & 0.19 m & N/A &  \cite{ma2020structured}{} \\ \hline
13 & Scalar time reversal & 0.3 m & N/A &  \cite{morales2015delivery}{} \\ \hline
14 & Scalar time reversal & 2 m & N/A &  \cite{czarske2016transmission}{} \\ \hline
15 & Mode-group excitation & 44.3 km & 6 mode groups &  \cite{ryf2015mode}{} \\ \hline
16 & Mode-group excitation & 5 km & 8 mode groups &  \cite{franz2012experimental}{} \\ \hline
17 & Mode-group excitation & 2.6 km & 4 mode groups &  \cite{zhu2017orbital}{} \\ \hline
\end{tabular}
\caption{Summary of methods for modal crosstalk suppression in standard multimode fibers. It can be seen that transfer matrix inversion has only been applied to short fibers that are less than 5-m long. This is because the characterization of a high-dimensional transfer matrix can take as long as hours. Since it is technically challenging to stabilize long fibers for hours, short fibers are used for experimental demonstrations. It can also be seen that scalar time reversal has only been used for short fibers. In \ref{sec:ScalarTimeReversal} we show that scalar time reversal cannot be used for a 1-km-long fiber due to the inevitable polarization mixing in long fibers. Mode-group excitation can be used to transmit mode groups over a long fiber. However, a standard MMF only supports a few mode groups, and thus this method is not comparable to our approach. HG: Hermite-Gauss. LG: Laguerre-Gauss. LP: linearly polarized.}
\end{table}

\clearpage
\section{--- Experimental setup}

\begin{figure}[h]
\center
\includegraphics[width=  1\linewidth]{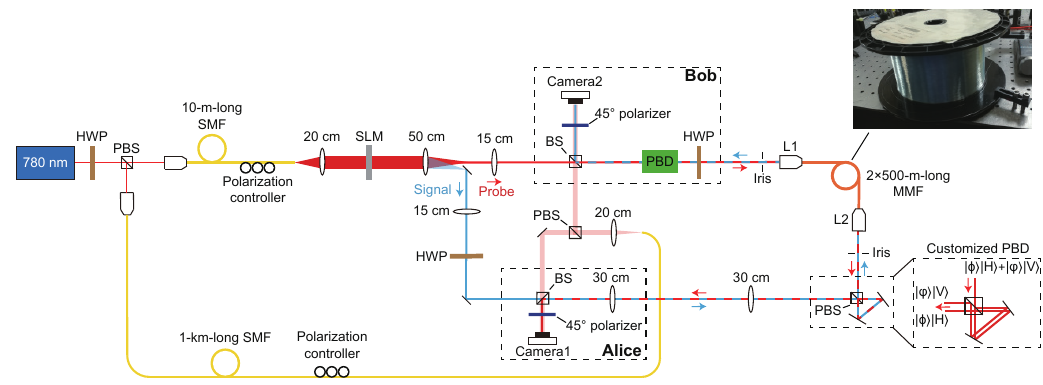}
\caption{The schematic of the experimental setup. HWP: half-wave plate. PBS: polarizing beamsplitter. SMF: single-mode fiber. SLM: spatial light modulator. BS: beamsplitter. PBD: polarizing beam displacer. MMF: multimode fiber. The same SLM is used to generate probe beams for Bob (denoted by red lines) and time-reversed signal beams (denoted by blue lines) for Alice by switching the phase grating. The multimode fiber spool is shown at the upper right corner and is resting on the optical table without any specialized thermal or mechanical stabilization. The optical table used in the experiment is not floated. The customized PBD is made of a PBS and two mirrors as illustrated by the inset. }
\label{fig:setup}
\end{figure}

The detailed experimental setup is shown in Supplementary Fig.~\ref{fig:setup}. A 780 nm laser (DL pro, Toptica) is used as the light source. The light is spatially filtered by a 10-m-long single-mode fiber (SMF) and then collimated to illuminate the spatial light modulator (SLM). A single SLM (Pluto 2 VIS-020, Holoeye) is used to generate both the probe beam for Bob and the signal beam for Alice with the choice being made by switching the overall phase grating written onto the SLM. A binary phase grating is used to generate complex-amplitude spatial modes in the first diffraction order \cite{mirhosseini2013rapid}{}. The probe beam generated by Bob is denoted by red lines and the signal beams generated by Alice is denoted by blue lines in Supplementary Fig.~\ref{fig:setup}. Bob uses a polarizing beam displacer (MBDA10, Karl Lambrecht) to generate a horizontally polarized spatial mode, and the polarization can be adjusted by a subsequent half-wave plate (HWP). The generated probe beam is then coupled into a 1-km-long multimode fiber (MMF) by an aspheric lens L1 (C110TMD-B, Thorlabs). The spatial mode beam waist size in the MMF used in our experiment is $w_0=5.06$ \textmu{}m \cite{renninger2013optical}{}. The MMF is comprised of two 500-m-long fibers (Clearcurve OM3, Corning) that are spliced together. The scrambled probe beam received by Alice is collimated by an aspheric lens L2 (C230TMD-B, Thorlabs), and a subsequent Sagnac interferometer is used as a customized polarizing beam displacer to coherently separate the horizontal and vertical polarization components of the scrambled probe beam to two beams that propagate along the same direction but are displaced with respect to each other \cite{perez2017demand}{}. The Sagnac interferometer as a customized polarizing beam displacer (PBD) provides more flexibility than the commercially available polarizing beam displacer because the transverse separation between the two displaced beams can be tuned by adjusting the mirrors in the Sagnac interferometer. A 1-km-long SMF is used to provide a coherent reference light source to interfere with the scrambled probe beam. A digital camera (Camera 1, BFS-U3-16S2M-CS, FLIR) is used by Alice to measure the interference pattern and perform the off-axis holography \cite{cuche2000spatial}{}. In this way, Alice can measure the amplitude, phase, and polarization of the scrambled probe beam via a single-shot measurement. The time-reversed signal beam is generated by the SLM and then directed back to the MMF. More details about the detection and generation of vector beam can be found in \cite{zhu2019single}{}, and the alignment procedure for time reversal can be found in \cite{jang2014method}{}. Bob measures the unscrambled signal beam by another camera (Camera 2, BFS-U3-31S4M-C, FLIR) and performs the digital spatial mode decomposition to obtain the crosstalk matrix.

\begin{figure}[t]
\center
\includegraphics[width=  1\linewidth]{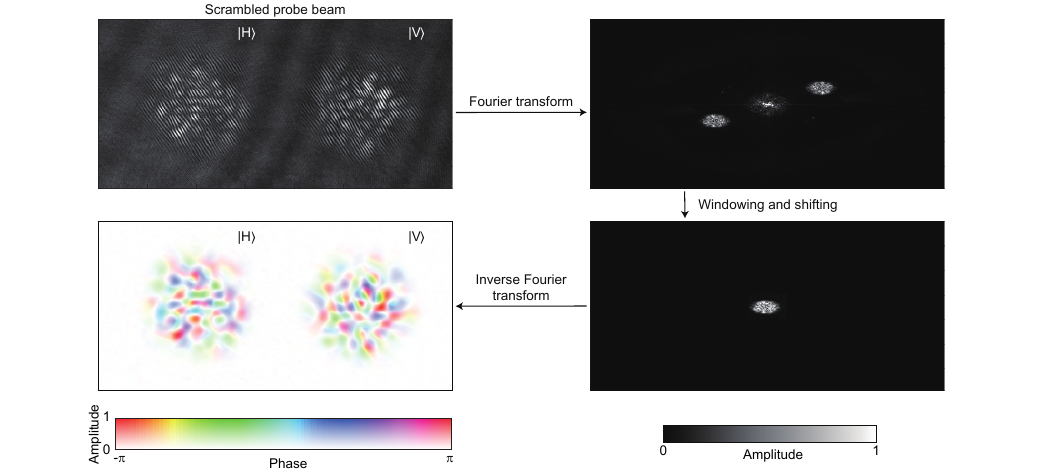}
\caption{Procedure for vectorial off-axis holography. The horizontally polarized beam and the vertically polarized beam are displaced by a PBD and then interfere with a 45\textdegree{} polarized beam (see Fig.~1 in the manuscript). The interference pattern is recorded by a camera. A Fourier transform is performed and the first-order component is selected and shifted to the image center. Then an inverse Fourier transform is performed to retrieve the polarization, amplitude, and phase of the scrambled probe beam. }
\label{fig:setup2}
\end{figure}

In our experiment, both Alice and Bob perform data processing for off-axis holography in MATLAB on a desktop computer (Intel i7-9700K with Nvidia RTX 2070 Super). The digital signal processing for off-axis holography involves several fast Fourier transforms (FFTs), which can be significantly sped up by using a dedicated digital signal processor. The procedure of vectorial off-axis holography is shown in Supplementary Fig.~\ref{fig:setup2}. The horizontally polarized component and the vertically polarized component interfere with a 45\textdegree{} polarized plane wave, and the interference pattern is recorded by a camera. Then a Fourier transform is performed, and the first-order component in the Fourier domain is selected and shifted to the center. Finally an inverse Fourier transform is performed, and thus the amplitude and phase of the scrambled probe beam are obtained. Ideally, the digital signal processing for Alice can even be avoided if the camera for off-axis holography and the SLM for generating the time-reversed signal beam are exactly placed at positions that are imaging planes with respect to each other \cite{jang2014method}. When the positions of camera and SLM are perfectly aligned, we can directly imprint the digital interference pattern recorded by the camera onto the SLM without performing any digital signal processing. Instead of carefully aligning the position of the camera, Alice digitally compensates the misalignments such as tip and tilt, transverse and axial displacement, and defocus (see \cite{jang2014method}). At Bob's side, we are using a second digital off-axis holography to measure the unscrambled signal beam, which allows us to obtain the crosstalk matrix with reduced experimental complexity. It should be noted that the off-axis holography can be readily replaced by a spatial mode sorter \cite{zhou2017sorting,zhou2018hermite, ruffato2018compact}{} to analyze the crosstalk matrix and enable high-speed spatial mode detection.

\clearpage

\section{--- Definition of the single mode index}
The Laguerre-Gauss and Hermite-Gauss mode are typically denoted by two indices, which are ($p$,$\ell$) for Laguerre-Gauss mode and ($m$,$n$) for Hermite-Gauss mode. For the convenience of presentation we use a single mode index $j=0,1,2,\cdots$ to denote the modes. Here we refer to the convention of Zernike polynomials \cite{thibos2002standards}{} and adopt the following definitions. For a specific mode index $j$, the mode group number $N$ can be calculated as $N=\text{ceil}((-3+\sqrt{9+8j})/2)$. Then we have $\ell=2j-N(N+2)$ and $p=(N-|\ell|)/2$ for Laguerre-Gauss mode. Given the one-to-one correspondence between Hermite-Gauss mode and Laguerre-Gauss mode \cite{beijersbergen1993astigmatic}{}, we can define $m=p+\text{max}(\ell,0)$ and $n=p-\text{min}(\ell,0)$ for Hermite-Gauss mode. On the other hand, for a given
($p$,$\ell$) or ($m$,$n$), we can calculate $j$ as $j=[N(N+2)+\ell]/2$ with $N=2p+|\ell|$ for Laguerre-Gauss modes and $j=[N(N+2)+m-n]/2$ with $N=m+n$ for Hermite-Gauss modes. Examples of the conversion relation between $j$, $N$, ($p$,$\ell$) and ($m$,$n$) are given in Supplementary Table.~\ref{tab:single}.

\begin{table}[h]
\center
\begin{tabular}{|c|c|c|c|c|c|c|}
\hline
$j$ & 0 & 1  & 2 & 3  & 4 & 5 \\ \hline
$N$ &0 & 1 & 1 & 2  & 2  & 2 \\ \hline
$p$ & 0 & 0  & 0 & 0  & 1 & 0 \\ \hline
$\ell$ & 0 & -1 & 1 & -2 & 0 & 2 \\ \hline
$m$ & 0 & 0  & 1 & 0  & 1 & 2 \\ \hline
$n$ & 0 & 1  & 0 & 2  & 1 & 0 \\ \hline
\end{tabular}
\caption{Examples of relation between the single mode index $j$, mode group number $N$, Laguerre-Gauss mode indices ($p,\ell$), and Hermite-Gauss mode indices ($m,n$).}
\label{tab:single}
\end{table}

\clearpage

\section{--- Crosstalk matrix and the normalized modal fidelity}
The 210$\times$210 unnormalized crosstalk matrix of the scrambled probe beams received by Alice when Bob transmits standard Laguerre-Gauss and Hermite-Gauss modes (i.e. in the absence of vectorial time reversal) are shown in Supplementary Fig.~\ref{fig:CMLGS} and Supplementary Fig.~\ref{fig:CMHGS}. Due to the strong spatial mode scrambling, the average unnormalized modal fidelity in this case is $\approx$1\% for both Laguerre-Gauss modes and Hermite-Gaussian modes. The 210$\times$210 unnormalized crosstalk matrix in the presence of vectorial time reversal are presented in Supplementary Fig.~\ref{fig:CMLG} and Supplementary Fig.~\ref{fig:CMHG}. The crosstalk matrix is calculated as follows. The received vectorial mode is denoted as $\ket{\phi}=\ket{\psi_1,\text{H}}+\ket{\psi_2,\text{V}}$, where H and V represent the horizontal and vertical polarization state, $\psi_{1}$ and $\psi_{2}$ represent the corresponding spatial mode, and $\ket{\phi}$ is normalized such that $\braket{\phi | \phi}=1$. Each element in the crosstalk matrix is the squared inner product between the received mode $\ket{\phi}$ and a particular Laguerre-Gauss or Hermite-Gauss mode. As an example, for a horizontally polarized Laguerre-Gauss mode $\ket{\text{LG}_j,\text{H}}$, the squared inner product can be expressed as $|\braket{\phi | \text{LG}_j,\text{H}}|^2$, where $0 \leq j \leq 104$ is the single mode index. For a Laguerre-Gauss mode, it is normalized such that $\braket{\text{LG}_j,\text{H} | \text{LG}_j,\text{H}}=1$ and $\braket{\text{LG}_j,\text{V} | \text{LG}_j,\text{V}}=1$. Similar normalization is also applied to Hermite-Gauss modes. It should be noted that since the spatial modes with $0 \leq j \leq 104$ do not form a complete basis set, the sum of each row in the crosstalk matrix is less than unity, $i.e.$ $\sum_{j=0}^{104} |\braket{\phi | \text{LG}_j,\text{H}}|^2 + |\braket{\phi | \text{LG}_j,\text{V}}|^2 <1$. However, we didn't normalize the crosstalk matrix and directly present the unnormalized modal fidelity in the manuscript. After normalizing the sum of each row of the crosstalk matrix to unity, the modal fidelity can be higher as shown in Supplementary Fig.~\ref{fig:NormFid}. It can be seen that the average of normalized modal fidelity is 91.5\% for Laguerre-Gauss modes and 89.3\% for Hermite-Gauss modes. This is because the crosstalk due to coupling to higher-order modes ($j\geq 105$) is discarded. This is permissible in an experiment because the higher-order modes can in principle be separated by a mode sorter in practical applications and thus does not contribute to the crosstalk.


To aid readers for analyzing the crosstalk matrix, we also present the crosstalk distributions in the following four categories. Here we assume the mode of interest is a horizontally polarized Laguerre-Gauss mode $\ket{\text{LG}_j,\text{H}}$ and the received time-reversed mode is $\ket{\phi}=\ket{\psi_1,\text{H}}+\ket{\psi_2,\text{V}}$ as an example. The four crosstalk categories are (1) crosstalk from coupling to modes inside the crosstalk matrix with degenerate polarization, which can be expressed as $C_1= \sum_k |\braket{\phi | \text{LG}_k,\text{H}}|^2$ for $0\leq k \leq 104$ and $k \neq j$. (2) crosstalk from coupling to modes inside the crosstalk matrix with orthogonal polarization, which can be expressed as $C_2= \sum_k |\braket{\phi | \text{LG}_k,\text{V}}|^2$ for $0\leq k \leq 104$. (3) crosstalk from coupling to modes outside the crosstalk matrix with degenerate polarization, which can be expressed as $C_3= \sum_k |\braket{\phi | \text{LG}_k,\text{H}}|^2$ for $ k \geq 105$ or equivalently $C_3= |\braket{\psi_1|\psi_1}|^2 -C_1 - |\braket{\psi_1| \text{LG}_j}|^2$. (4) crosstalk from coupling to modes outside the crosstalk matrix with orthogonal polarization, which can be expressed as $C_4= \sum_k |\braket{\phi | \text{LG}_k,\text{V}}|^2$ for $ k \geq 105$ or equivalently $C_4=|\braket{\psi_2|\psi_2}|^2 -C_2$. These results are shown in Supplementary Fig.~\ref{fig:FidAnalysis}.

\begin{figure}[t]
\center
\includegraphics[width=  1\linewidth]{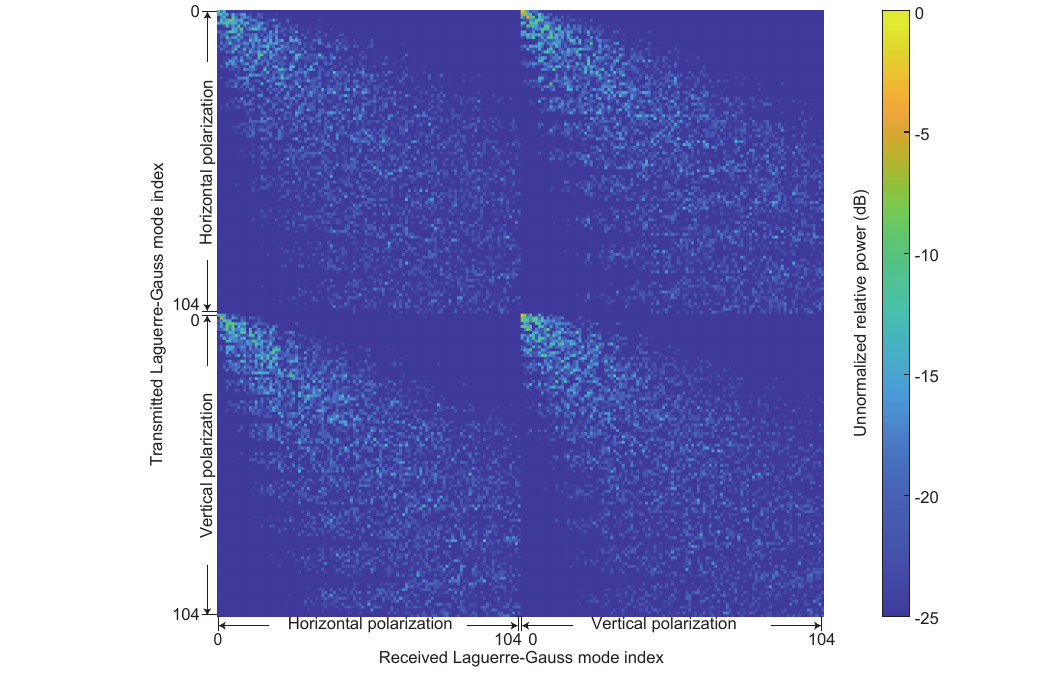}
\caption{Unnormalized 210$\times$210 crosstalk matrix on a logarithmic scale for scrambled probe beams received by Alice when Bob transmits Laguerre-Gauss modes in the absence of vectorial time reversal.}
\label{fig:CMLGS}
\end{figure}

\begin{figure}[t]
\center
\includegraphics[width=  1\linewidth]{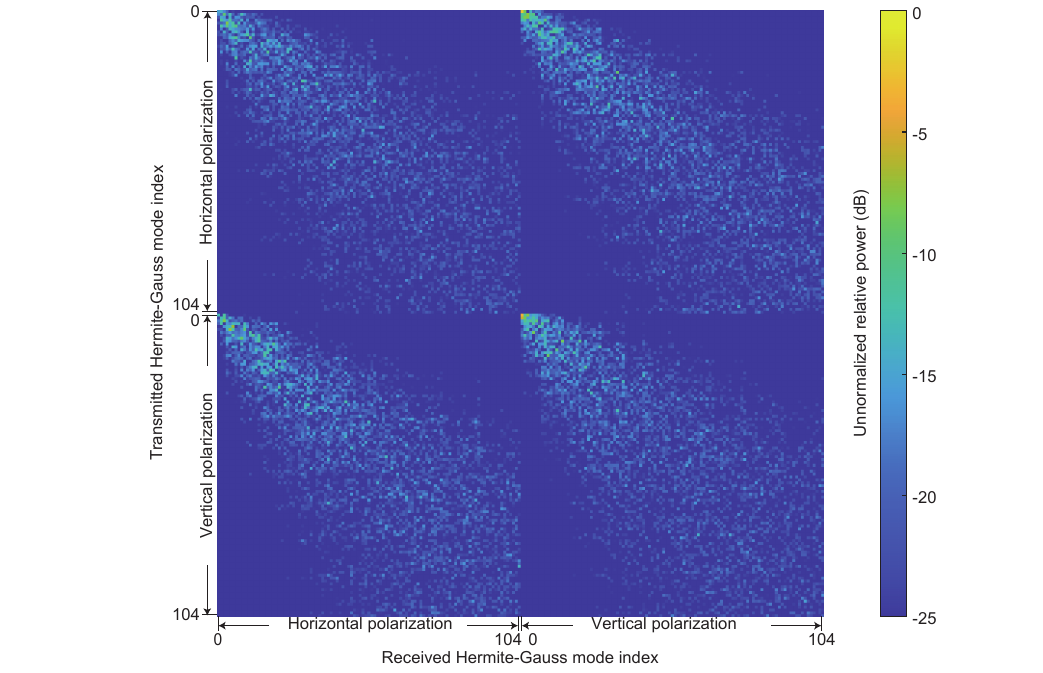}
\caption{Unnormalized 210$\times$210 crosstalk matrix on a logarithmic scale for scrambled probe beams received by Alice when Bob transmits Hermite-Gauss modes in the absence of vectorial time reversal.}
\label{fig:CMHGS}
\end{figure}

\begin{figure}[t]
\center
\includegraphics[width=  1\linewidth]{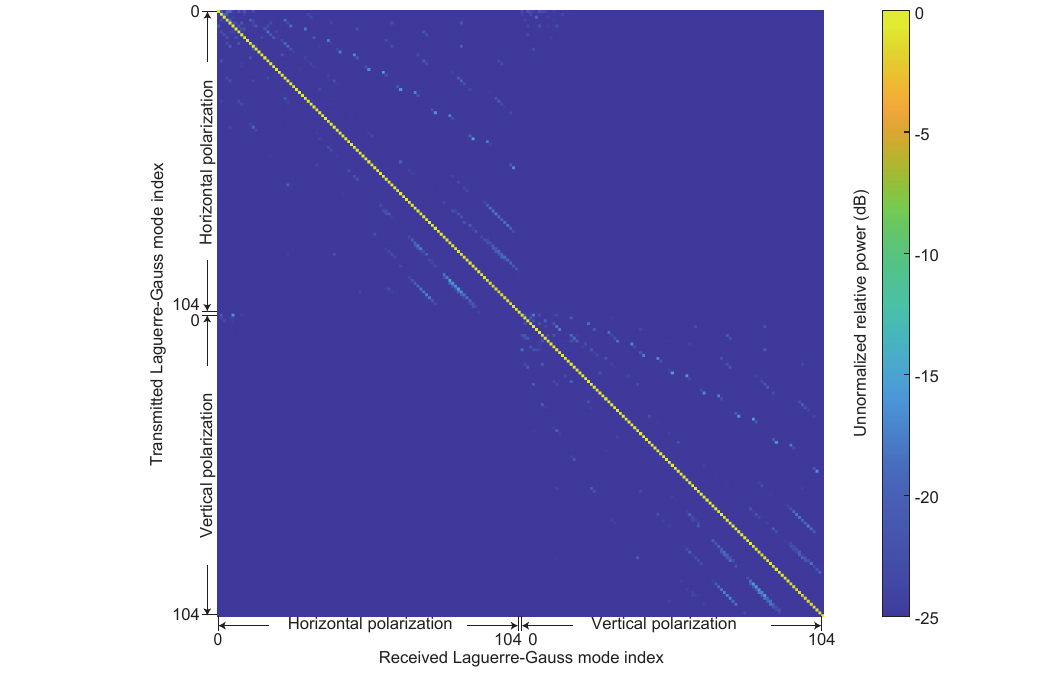}
\caption{Unnormalized 210$\times$210 crosstalk matrix on a logarithmic scale for unscrambled Laguerre-Gauss modes received by Bob when performing vectorial time reversal.}
\label{fig:CMLG}
\end{figure}

\begin{figure}[t]
\center
\includegraphics[width=  1\linewidth]{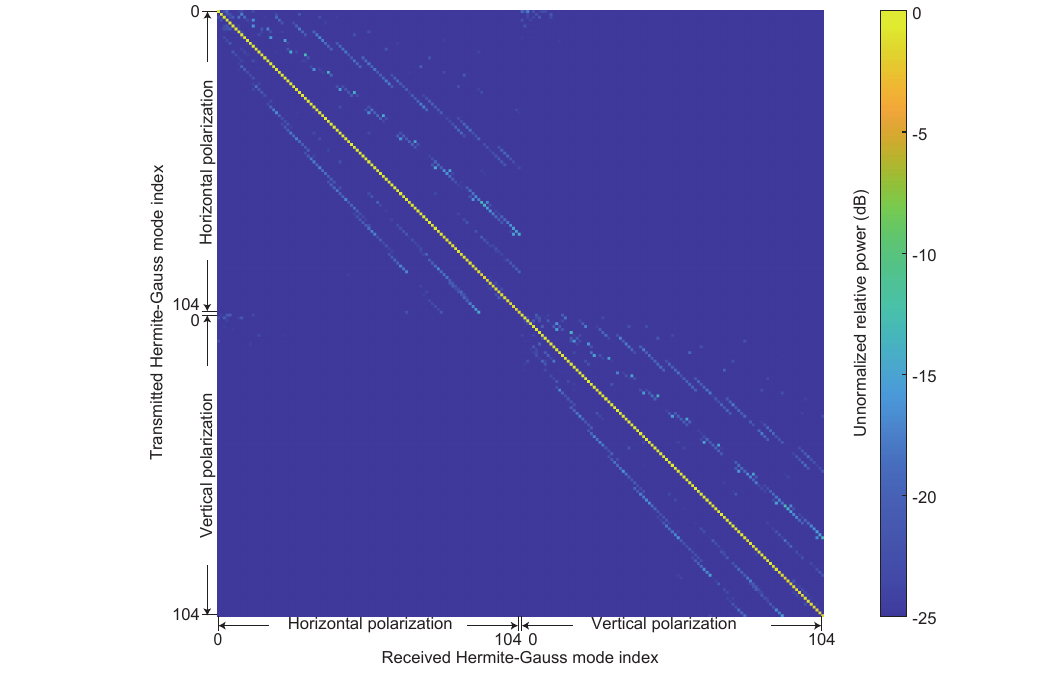}
\caption{Unnormalized 210$\times$210 crosstalk matrix on a logarithmic scale for unscrambled Hermite-Gauss modes received by Bob when performing vectorial time reversal.}
\label{fig:CMHG}
\end{figure}

\clearpage
\begin{figure}[t]
\center
\includegraphics[width=  1\linewidth]{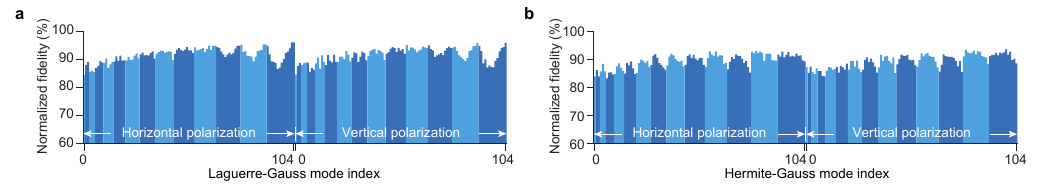}
\caption{Normalized modal fidelity for unscrambled (\textbf{a}) Laguerre-Gauss and (\textbf{b}) Hermite-Gauss modes. The normalized modal fidelity is calculated by dividing the corresponding diagonal element (i.e. the unnormalized modal fidelity) of the crosstalk matrices (shown in Supplementary Fig.~\ref{fig:CMLG} for Laguerre-Gauss modes and Supplementary Fig.~\ref{fig:CMHG} for Hermite-Gauss modes) by the sum of elements in each column.}
\label{fig:NormFid}
\end{figure}

\begin{figure}[h]
\center
\includegraphics[width=  1\linewidth]{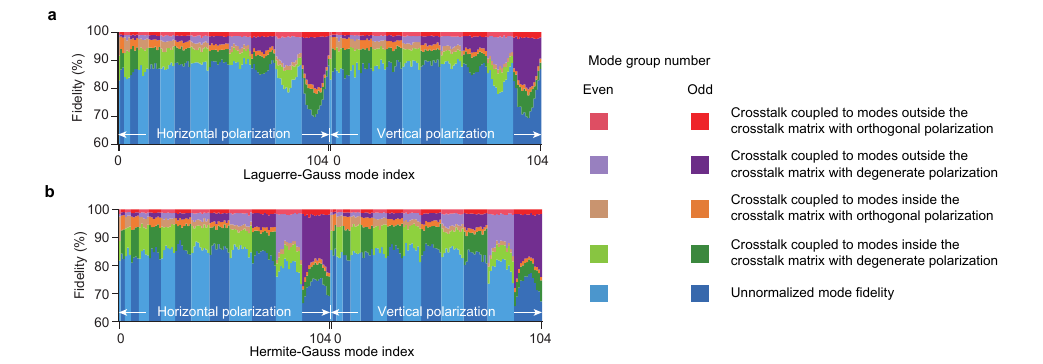}
\caption{The modal fidelity and crosstalk distributions for unscrambled (\textbf{a}) Laguerre-Gauss modes and (\textbf{b}) Hermite-Gauss modes.  }
\label{fig:FidAnalysis}
\end{figure}

\clearpage
\section{--- Experimental generation fidelity of SLM}

In the manuscript we show that the average modal fidelity without normalization is 85.6\% for Laguerre-Gauss modes and 82.6\% for Hermite-Gauss modes. The main reason for imperfect modal fidelity is attributed to the imperfect spatial mode generation fidelity of the SLM. In the experiment, Alice uses a SLM to generate the phase conjugate of the scrambled probe beam, and Bob uses a SLM to generate the Laguerre-Gauss or Hermite-Gauss modes. We characterize the fidelity of the generated spatial modes and horizontal polarization component of scrambled probe beams using off-axis holography, with the results shown in Supplementary Fig.~\ref{fig:SLM}. We also measure the complex field of the generated signal beam $\ket{\psi_{s}}$, and calculate the overlap integral with the scrambled probe beam $\ket{\psi_{p}}$ to get the signal beam fidelity as $| \braket{ \psi_{s} | \psi_{p} }|^2$. We take the product of the probe spatial modal fidelity and the signal beam fidelity as the experimental generation fidelity, which is a simple estimate of time-reversed modal fidelity. In Supplementary Fig.~\ref{fig:SLM}(a-d) we show the ideal and generated spatial mode and time-reversed signal beam for HG(2,5) and LG(1,4) modes. In Supplementary Fig.~2(c,d) in the manuscript we present the experimental generation fidelity for individual Laguerre-Gauss and Hermite-Gauss modes, which shows reasonable agreement with the fidelity of time-reversed modes. In general, the modal fidelity for Laguerre-Gauss modes is slightly higher than that of Hermite-Gauss modes, and the Laguerre-Gauss modes with $\ell \neq 0$ mode have a higher fidelity than those with $\ell=0$. In addition, the signal beam generation has a slightly lower fidelity than the well-defined Laguerre-Gauss and Hermite-Gauss modes. The maximum experimental generation fidelity we measured is 90.8\% for Laguerre-Gauss modes and 88.4\% for Hermite-Gauss modes, and the average experimental generation fidelity is 85.8\% for Laguerre-Gauss modes and 83.7\% for Hermite-Gauss modes. Thus, by improving the experimental generation fidelity through using a higher-quality SLM, the performance of vectorial time reversal can be further enhanced.

\begin{figure}[h]
\center
\includegraphics[width=  1\linewidth]{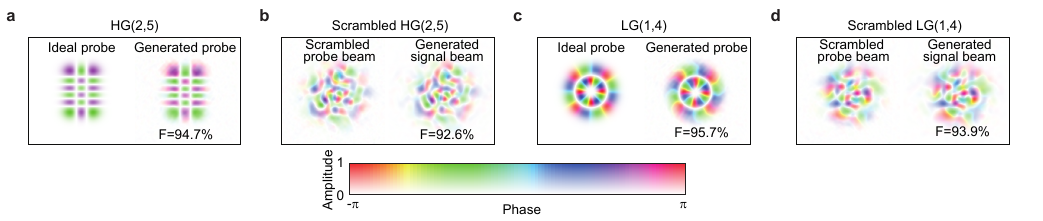}
\caption{(\textbf{a}--\textbf{d}) The mode generation fidelity for HG(2,5) and LG(1,4) modes. The HG(2,5) and LG(1,4) probe beams generated by SLM are shown at the right panel of (\textbf{a}) and (\textbf{c}), and the corresponding ideal modes are shown at the left panel for comparison. The corresponding phase conjugate of the scrambled probe beams are shown at the left panel of (\textbf{b}) and (\textbf{d}), and the generated signal beams are shown at the right panel. The calculated fidelity (F) is listed below the generated modes. }
\label{fig:SLM}
\end{figure}

\clearpage
\section{--- Experimental results of scalar time reversal}\label{sec:ScalarTimeReversal}
Here we show the results for scalar time reversal. Bob transmits horizontally polarized spatial modes to Alice. While Alice can measure both the horizontal and vertical polarization components of the scrambled probe beam, she only generates the phase conjugate of the horizontal polarization and ignores the vertical polarization. The experimental results are shown in Supplementary Fig.~\ref{fig:scalar}. It can be seen that the modal fidelity of scalar time reversal is significantly worse than that of vectorial time reversal. The average modal fidelity is 41.2\% for Laguerre-Gaussian modes and 39.7\% for Hermite-Gauss modes. The results here confirm the necessity of vectorial time reversal.

\begin{figure}[h]
\center
\includegraphics[width=  1\linewidth]{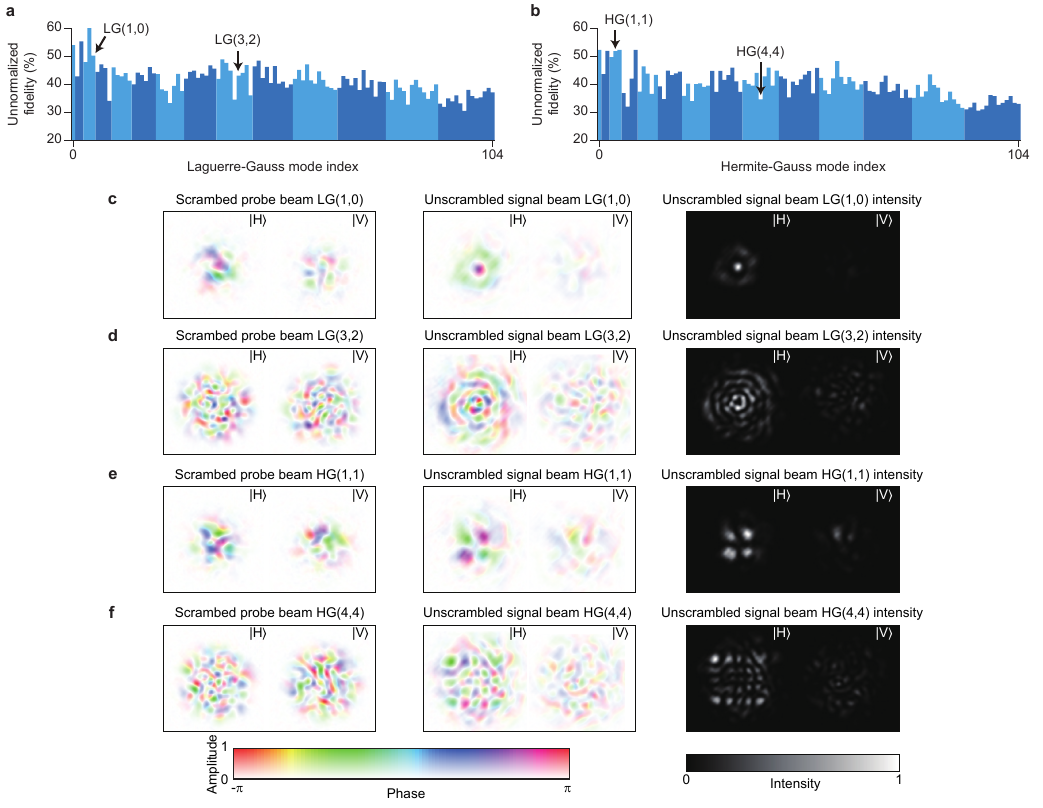}
\caption{(\textbf{a}, \textbf{b}) The measured modal fidelity for unscrambled Laguerre-Gauss modes and Hermite-Gauss modes received by Bob when Alice performs scalar time reversal (i.e. using one polarization only). (\textbf{c}--\textbf{f}) The measured scrambled probe beam, the unscrambled signal beam, and the corresponding intensity for LG(1,0), LG(3,2), HG(1,1), HG(4,4) modes respectively.   }
\label{fig:scalar}
\end{figure}

\clearpage
\section{--- Polarization crosstalk matrix}
In the manuscript we show the normalized polarization crosstalk matrix within the individual spatial mode subspace (see Fig. 3a in the manuscript). Here we continue to use LG$_j$ to show how the calculation is performed. For the received time-reversed mode $\ket{\phi}$, we first calculate the unnormalized squared inner product for horizontally (H), vertically (V), diagonally (D), and anti-diagonally (A) polarized LG$_j$ mode as $F_H^u=|\braket{\phi | \text{LG}_j,\text{H}}|^2$, $F_V^u=|\braket{\phi | \text{LG}_j,\text{V}}|^2$, $F_D^u=|\braket{\phi | \text{LG}_j,\text{D}}|^2$, and $F_A^u=|\braket{\phi | \text{LG}_j,\text{A}}|^2$, where $\ket{\text{D}}=(\ket{\text{H}}+\ket{\text{V}})/\sqrt{2}$ and $\ket{\text{A}}=(\ket{\text{H}}-\ket{\text{V}})/\sqrt{2}$. Then the normalized fidelity can be calculated as $F_{\text{H}}=F_\text{H}^\text{u}/(F_\text{H}^\text{u}+F_\text{V}^\text{u})$, $F_\text{V}=F_\text{V}^\text{u}/(F_\text{H}^\text{u}+F_\text{V}^\text{u})$, $F_\text{D}=F_\text{D}^\text{u}/(F_\text{D}^\text{u}+F_\text{A}^\text{u})$, and $F_\text{A}=F_\text{A}^\text{u}/(F_\text{D}^\text{u}+F_\text{A}^\text{u})$. These four numbers form one row of the crosstalk matrix, and the rest of the matrix can be similarly calculated. The normalized polarization crosstalk matrix for Hermite-Gauss modes is also calculated in this way. Since the calculation is performed for a (4$\times$4)-dimensional state space, the crosstalk is small. It should be noted that such calculation is permissible because in an experiment one can use a polarization-independent spatial mode sorter and a polarizing beamsplitter to access the (4$\times$4)-dimensional state space, and polarization crosstalk that couples to other spatial modes can be discarded.

\end{document}